\documentclass[aps,prl,preprint,superscriptaddress]{revtex4-1}
\newcommand{\figSize}{1}
\newcommand{\figSizeB}{0.6}

\usepackage[english]{babel}
\usepackage{graphicx}
\usepackage{natbib}
\usepackage{color}

\makeatletter
\@ifundefined{textcolor}{}
{%
 \definecolor{BLACK}{gray}{0}
 \definecolor{WHITE}{gray}{1}
 \definecolor{RED}{rgb}{1,0,0}
 \definecolor{GREEN}{rgb}{0,1,0}
 \definecolor{BLUE}{rgb}{0,0,1}
 \definecolor{CYAN}{cmyk}{1,0,0,0}
 \definecolor{MAGENTA}{cmyk}{0,1,0,0}
 \definecolor{YELLOW}{cmyk}{0,0,1,0}
 }

\begin{document}
\title{Strong interactions between dipolar polaritons}
\author{Emre Togan$^\dagger$}
\affiliation{Institute of Quantum Electronics, ETH Zurich, CH-8093
Zurich, Switzerland.}

\author{Hyang-Tag Lim$^\dagger$}
\affiliation{Institute of Quantum Electronics, ETH Zurich, CH-8093
Zurich, Switzerland.}

\author{Stefan Faelt}
\affiliation{Institute of Quantum Electronics, ETH Zurich, CH-8093
Zurich, Switzerland.}
\affiliation{Solid State Physics Laboratory, ETH Zurich, CH-8093 Zurich, Switzerland\\
$^\dagger$These authors contributed equally to this work.\\
$^*$Correspondence and requests for materials should be addressed to A.I. (imamoglu@phys.ethz.ch)}

\author{Werner Wegscheider}
\affiliation{Solid State Physics Laboratory, ETH Zurich, CH-8093 Zurich, Switzerland\\
$^\dagger$These authors contributed equally to this work.\\
$^*$Correspondence and requests for materials should be addressed to A.I. (imamoglu@phys.ethz.ch)}

\author{Atac Imamoglu$^*$}
\affiliation{Institute of Quantum Electronics, ETH Zurich, CH-8093
Zurich, Switzerland.}


\begin{abstract}
\textbf{
Nonperturbative coupling between cavity photons and excitons leads to formation of hybrid light-matter excitations termed polaritons. In structures where photon absorption leads to creation of excitons with aligned permanent dipoles~\cite{cristofolini_coupling_2012,rosenberg_electrically_2016,rosenberg_strongly_2018}, the elementary excitations, termed dipolar polaritons, are expected to exhibit enhanced interactions~\cite{byrnes_effective_2014,nalitov_voltage_2014}. Here, we report a substantial increase in interaction strength between dipolar polaritons as the size of the dipole is increased by tuning the applied gate voltage. To this end, we use coupled quantum well structures embedded inside a microcavity where coherent electron tunneling between the wells controls the size of the excitonic dipole.  Modifications of the interaction strength are characterized by measuring the changes in the reflected intensity of light when polaritons are driven with a resonant laser.  Factor of 6.5 increase in the interaction strength to linewidth ratio that we obtain indicates that dipolar polaritons could be used to demonstrate a polariton blockade effect~\cite{kyriienko_tunable_2014} and thereby form the building blocks of many-body states of light~\cite{carusotto_quantum_2013}.
}
\end{abstract}

\maketitle

Realization of strongly interacting photonic systems is one of the holy grails of quantum optics. Substantial progress towards this goal has been achieved using Rydberg polaritons -- quasiparticles consisting of a propagating photon and a collective Rydberg excitation: van der Waals interactions between Rydberg atoms ensures that the polaritonic excitations interact strongly~\cite{pritchard_cooperative_2010,peyronel_quantum_2012,dudin_strongly_2012,tiarks_single-photon_2014,gorniaczyk_single-photon_2014,jia_strongly_2018}. 
In solid-state cavity-polariton systems consisting of a cavity photon and a quantum well exciton, dominant direct exciton (DXs) interactions originate from short-range exchange terms ~\cite{ciuti_role_1998,tassone_exciton-exciton_1999}. These interactions have lead to manifestation of a number of intriguing collective phenomena ranging from formation of spontaneous coherence~\cite{kasprzak_boseeinstein_2006}, through observation of vortex-antivortex pairs~\cite{nardin_hydrodynamic_2011,sanvitto_all-optical_2011} and dark solitons~\cite{amo_polariton_2011,grosso_soliton_2011}, to realization of polariton Josephson effect~\cite{lagoudakis_coherent_2010,abbarchi_macroscopic_2013}. However, a mean field approach could be used to accurately describe all these observations. Very recently, photon correlation measurements on strongly confined polaritons have demonstrated weak quantum correlations \cite{munoz-matutano_quantum-correlated_2017, delteil_preparation_nodate}. 

Increasing polariton-polariton interaction further is crucial to explore physics beyond mean-field and to explore a new regime of strongly correlated photons.  One way to enhance interactions is to engineer polaritonic excitations with a permanent dipole moment~\cite{cristofolini_coupling_2012}: such dipolar polaritons emerge as elementary optical excitations when DXs in a quantum well (QW) are strongly coupled to both microcavity photons and indirect dipolar excitons. A number of  studies have shown that interactions between polaritons can be enhanced by increasing the size of the optically induced dipole in  a wide QW~\cite{rosenberg_electrically_2016,rosenberg_strongly_2018}, or alternatively, the indirect exciton (IX) content~\cite{byrnes_effective_2014,nalitov_voltage_2014}. The structure we employ in our experiment allows us to tune the IX content and to increase the ratio of the interaction strength of polaritons to their linewidth without substantially compromising the exciton-photon coupling strength. As we demonstrate in this work, the associated dipole-dipole interactions can be much stronger than the intrinsic interactions between DXs and thereby provide a promising platform to realize many-body states of photons~\cite{carusotto_quantum_2013}.

\begin{figure}[p]
\includegraphics[width= \figSize \textwidth]{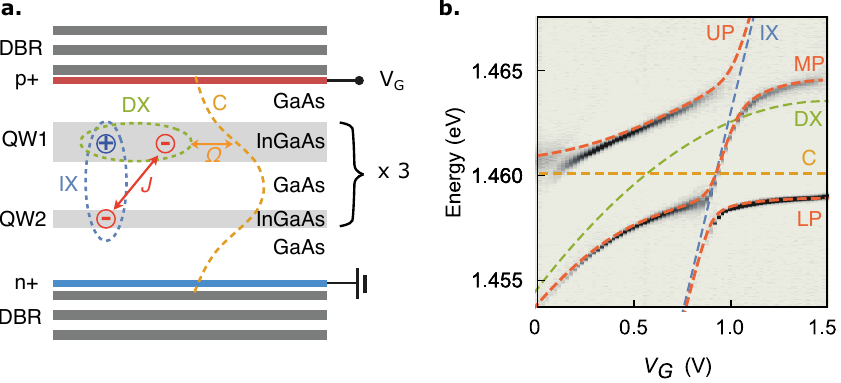}
\caption{ \textbf{Schematic of the sample structure and gate dependent reflection spectrum.} \textbf{a.} The sample contains three pairs of In$_{0.04}$Ga$_{0.96}$As coupled quantum wells (QWs) located at three separate antinodes of a $5\lambda/2$ cavity formed between two GaAs/AlAs distributed Bragg reflectors (DBRs). To ensure electron tunneling for lowest energy electron states of the two QWs occur at finite electric field we use QWs with different thicknesses, thus each pair consists of a 5 nm and a 11 nm thick QWs. The thickness of GaAs barrier between two QWs is 13 nm. An electric potential $V_\mathrm{G}$ applied between the $p$ (66 nm) and $n$ (66 nm) doped layers tunes the energy of direct exciton (DX) and indirect exciton (IX) levels.  See Supplementary Note 1 for details on the sample structure and the fabrication process. \textbf{b.} Reflection spectrum with zero in-plane momentum excitation ($\mathbf{k_{\rm in}} = 0 $) as a function of applied gate voltage $V_\mathrm{G}$. Changes in the energy levels of the bare DX (green dashed line) and IX (blue dashed lines) are extracted from a measurement of the reflection spectrum at a point where the top DBR has been etched (see Supplementary Figure 2). Cavity resonance (orange dashed line) is constant as we vary $V_{\rm G}$. Changes of the energies of the three coupled states, upper polariton (UP), middle polariton (MP), and lower polartion (LP),  with $V_\mathrm{G}$ is shown as red dashed lines. We use the calculated eigenstates associated with the red lines to estimate the DX, IX, and C (cavity) content at a particular sample position and $V_\mathrm{G}$. 
}
\label{fig1}
\end{figure}


We work with a microcavity sample, illustrated in Figure~\ref{fig1}\textbf{a},  which contains three pairs of coupled In$_{0.04}$Ga$_{0.96}$As QWs embedded inside a $p$-$i$-$n$ diode. By applying a voltage ($V_{\mathrm{G}}$) between the $p$ and $n$ doped regions we control the electric field $E_z$ in the growth direction and adjust the detuning between different QW exciton states. A DX state localized on the thick (11 nm) QW experiences quantum confined Stark shift and its energy decreases quadratically with the applied electric field. The same electric field leads to a linear shift in energy for an IX state with the hole localized on the thick QW and the electron on the thin (5 nm) QW. Separated by a thin (13 nm) GaAs layer, proximity of the two QWs facilitates electron tunneling with rate $J = 3.5$ meV between the two wells. When the energy of the DX state and the IX state are equal, this tunneling hybridizes the two exciton states. Due to its large oscillator strength DX state also couples strongly to the cavity mode localized between the two distributed Bragg reflectors (DBRs) at a rate $\Omega$.  The Hamiltonian  of the coupled system for zero in-plane wavevector $\mathbf{k_{\rm in}} = 0$ can be written in the matrix form
\begin{eqnarray}
H = \left(\begin{array}{c c c}
\epsilon_{\mathrm{IX}} -e d E & J/2 & 0 \\
J/2 & \epsilon_{\mathrm{DX}} - \alpha E^2 & \Omega/2 \\
0 & \Omega/2 & \epsilon_{\mathrm{C}}
 \end{array} \right),
\label{eqnCoupling}
\end{eqnarray}
where $e$ is the elementary charge, $d$ is the IX dipole size, $\alpha$ is the polarizability of DX, $\epsilon_{\mathrm{DX}}$ and $\epsilon_{\mathrm{IX}}$ are zero electric field energies of the DX and IX states, and $\epsilon_{\mathrm{C}}$ is the energy of the cavity mode at $\mathbf{k_{\rm in}} = 0$. We assume that the coupling between IX and cavity mode is negligible. The quasiparticle composition of the eigenstates, named lower, middle and upper polariton modes are characterized by generalized Hopfield coefficients $x$, $y$ and $c$ so that an eigenstate can be written as : $x \left| \mathrm{DX} \right \rangle + y \left| \mathrm{IX} \right \rangle + c \left| \mathrm{C} \right \rangle$ where $|x|^2 + |y|^2 + |c|^2= 1$.

We probe the sample held at 4 K by focusing our illumination beam down to a $\sim 5 \, \mu$m spot and observing changes in the reflection of the illumination beam. As shown in Figure~\ref{fig1}\textbf{b}, the reflection spectrum is dominated by three modes that we identify as the three polariton modes. Energies of the three polariton modes agree well with the eigenvalues of the matrix in Equation~\ref{eqnCoupling} with $d = 21$ nm, $\alpha = 1.2 \times 10^{-15}$ eV-m$^2$-V$^{-2}$, $\epsilon_{\mathrm{IX}} = 1.48209$ eV, $\epsilon_{\mathrm{DX}} = 1.46354$ eV, $\epsilon_{\mathrm{C}} = 1.4601$ eV, and $\Omega = 4.6$ meV. We use coefficients $x$, $y$ and $c$ that we extract from this model to identify the  particle weight for different polariton modes as a function of $V_{\mathrm{G}}$. We will focus on the lower polariton for the rest of this work.

\begin{figure}[p]
\includegraphics[width = \figSize \textwidth]{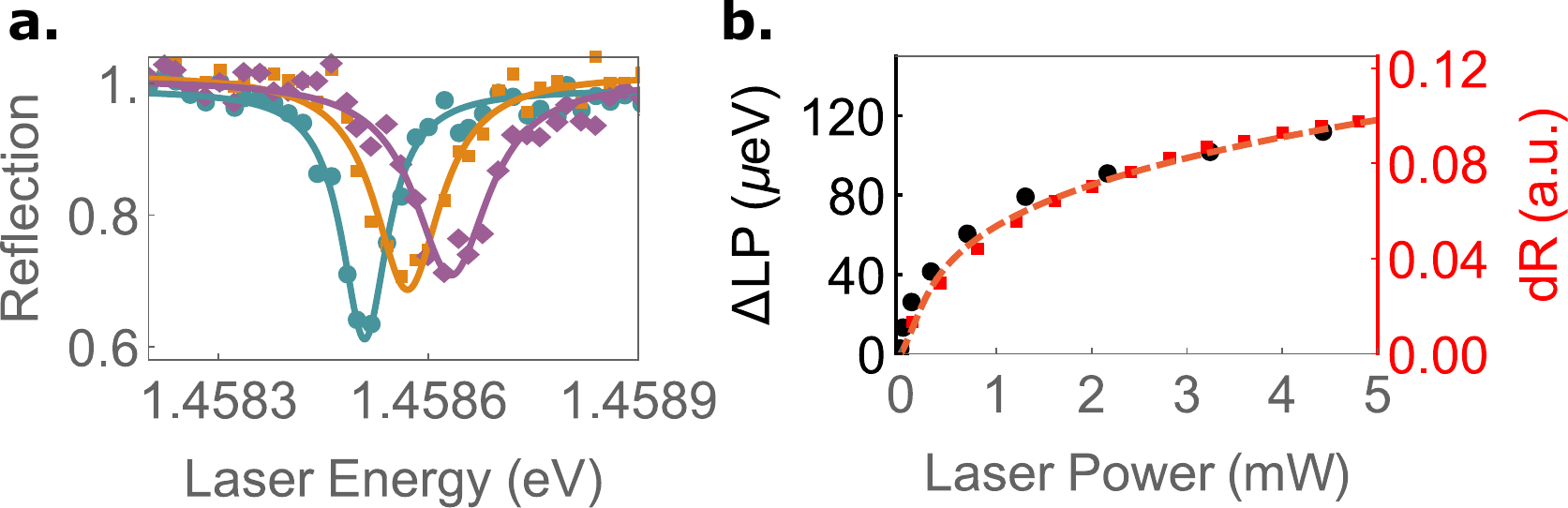}
\caption{ \textbf{Intensity dependent saturation of blue-shift of lower polaritons and of differential reflection at $V_{\mathrm G} = 1.1$ V.} \textbf{a.} Reflection spectrum obtained when scanning a weak ($\sim$ 0.5 $\mu$W) linearly polarized laser across a LP resonance. The power of a second, orthogonally linearly polarized laser tuned to 1.45853 eV, laser is varied (cyan : 0 mW, orange: 2.15 mW, purple: 7.2 mW). Polariton linewidth broadens as we increase drive power: $\gamma_{\rm P}$ are 80 $\mu$eV, 116 $\mu$eV, and 130 $\mu$eV, respectively. \textbf{b.} The left axis shows changes in the LP resonance energy, extracted from traces as shown in \textbf{a}, as the power of a second laser (horizontal axis) is varied. The right axis shows changes in the differential reflection of the laser tuned to 1.45853 eV as its power is varied. Differential reflection ($dR$) is obtained by subtracting the reflection signal obtained at $V_{\mathrm G} = 1.1$ V from a reflection trace obtained at $V_{\mathrm G} = 0 $ V. Red dashed line is the calculated $dR$ signal using Equation 2. The excitation laser is off-resonant with all polariton transitions at $V_{\mathrm G} = 0$ V. All reflection and two laser experiments are carried out using a pulse scheme to minimize effects of light induced slow changes to the polariton environment (see Supplementary Note 3).
}
\label{FigSaturation}
\end{figure}

Figure~\ref{FigSaturation}\textbf{a} illustrates the reflection spectrum obtained when the frequency of a low intensity ($\sim$ 0.5 $\mu$W) linearly polarized laser is scanned across a lower polariton resonance with $|x|^2 = 0.27$, $|y|^2 = 0.01$ and $|c|^2 = 0.72$ ($V_{\rm G} = 1.1 $ V). The polariton resonance blue shifts as the intensity of a second, orthogonally polarized, laser tuned to the polariton resonance (1.45853 meV) is increased. The blue shift, which is initially linear with the intensity of the second laser, saturates for high intensities (Figure~\ref{FigSaturation}\textbf{b}). Intensity of the differential reflection from the sample also exhibits saturation behavior with power (Figure~\ref{FigSaturation}{\textbf{b}}).  We use this saturation behavior to characterize the nonlinearity of the polariton mode.

\begin{figure}[p]
\includegraphics[width = \figSize \textwidth]{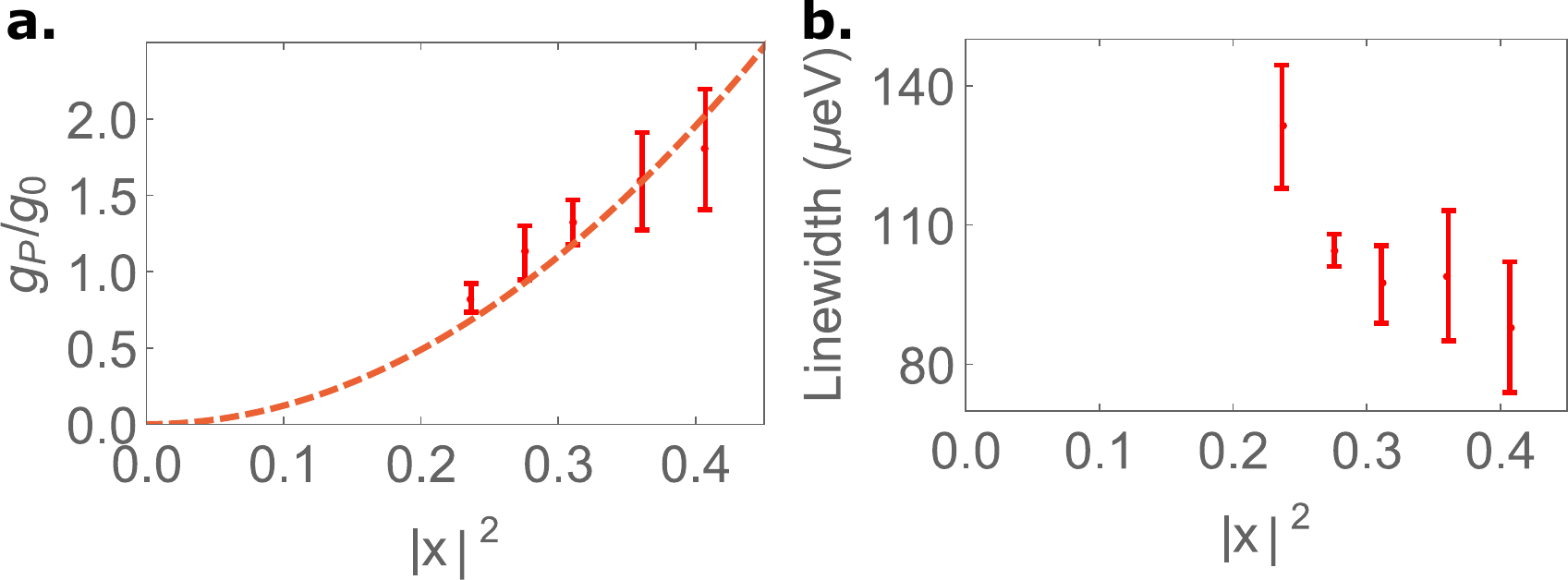}
\caption{ \textbf{Increase of polariton interactions with increasing DX content.} \textbf{a.} Changes in polariton interaction strength $g_{\mathrm{P}}$ extracted from differential reflection measurements similar to Figure~\ref{FigSaturation}\textbf{b} as a function of $|x|^2$. Different data points were obtained at $V_{\rm G}= 1.4 $ V at different positions on the sample. The IX content remains $|y|^2 \leq 0.004$ for all of the data (largest at high DX content). Changes in $g_{\rm P}$ are in agreement with $g_\mathrm{P} = (12 \pm 1) |x|^4 g_0$. Errorbars are standard deviation of the extracted $g_{\mathrm P}$ obtained at different repetition of the experiment at different spatial positions that have the same $|x|^2$ content (See Supplementary Note 4). \textbf{b.} Changes in the linewidth extracted from a low power ($\leq 0.5 $ $\mu$W) reflection measurements using a single laser. Errorbars are the standard deviation of the extracted linewidth when repeating the experiment at different spatial positions that have the same $|x|^2$ content.
}
\label{FigDXcontent}
\end{figure}

To quantitatively extract the strength of the nonlinearity we model the polariton behaviour, in the mean field limit, by a single nonlinear mode with a Kerr like nonlinearity~\cite{baas_optical_2004}.  Within this approximation, the equation of motion for the mean field expectation value of the lower polariton annihilation operator $p$ can be written as:

\begin{equation}
\frac{dp}{dt} = -\left(\frac{\gamma_{\rm P}}{2}+i\delta_{\rm P} \right) p -i g_{\rm P} \left| p \right|^2 p -c \sqrt{\gamma_{{\rm in}}} \sqrt{I^{{\rm in}}},
\label{0dGP}
\end{equation}
where  $\gamma_{\rm P}= 103$ $\mu$eV is the position averaged full width at half maximum (FWHM) polariton linewidth (see Figure \ref{FigDXcontent}{\textbf b}), $\delta_{\rm P} =\epsilon_{\mathrm{pol}}-\epsilon_{\mathrm{laser}}$ is the detuning of the laser from the polariton mode, $\gamma_{{\rm in}} = \gamma_{\mathrm{P}}/6$ characterizes the input coupling rate, and $I^{{\rm in}}$ characterizes the input/drive intensity. The nonlinear coefficient of the effective single-mode is $g_{\rm P} = U/A$, where $U$ is the strength of the two-dimensional (2D) polariton interaction and $A\sim 7 \,\mu$m$\times7\,\mu$m denotes the mode area; the latter should ideally be determined by solving the Gross Pitaevski equation (GPE) in the 2D limit (see Supplementary Note 6). With $\delta_{\rm P} = 0 $ and in steady state an increase in the polariton population leads to a blue-shift of the polariton resonance, hence as $I^{{\rm in}}$ increases the blue shift leads to a saturation of the polariton population, as well as a saturation of the differential reflection signal. For each position and gate voltage we fit a value of $g_{\mathrm P}$ so that the calculated differential reflection signal matches the measured differential reflection signal (see Supplementary Note 2).

In order to demonstrate the utility and limitations of this method in characterizing the strength of the polariton nonlinearity, we measure changes of $g_{\rm P}$ for a fixed $V_{\rm G} = 1.4 $ V as we vary cavity and DX contents. At this gate voltage the IX content as well as the induced dipole moment of DX is negligible. We use the wedge that is present in the sample which leads to position dependent changes in $\epsilon_{\mathrm{C}}$ to vary the DX and cavity contents of lower polaritons. Figure~\ref{FigDXcontent}\textbf{a} illustrates the  changes in the extracted $g_{\rm P}$ as a function of $|x|^2$ measured at different positions on the sample. The data is in agreement with $g_{\rm P} = 12 |x|^4 g_0$,  with $g_{0} = 1.6$ neV. This $|x|^4$ dependence is the expected behavior for DX induced polariton-polariton interactions~\cite{carusotto_quantum_2013}. We note that the standard deviation of the extracted $g_{\mathrm P}$ and polariton linewidth are large. These large variations in the extracted parameter are caused by residual effects of a position dependent slow charge environment that leads to sub linewidth shifts of the polariton resonance (see Supplementary Note 4).

\begin{figure}[p]
\includegraphics[width =\figSizeB  \textwidth]{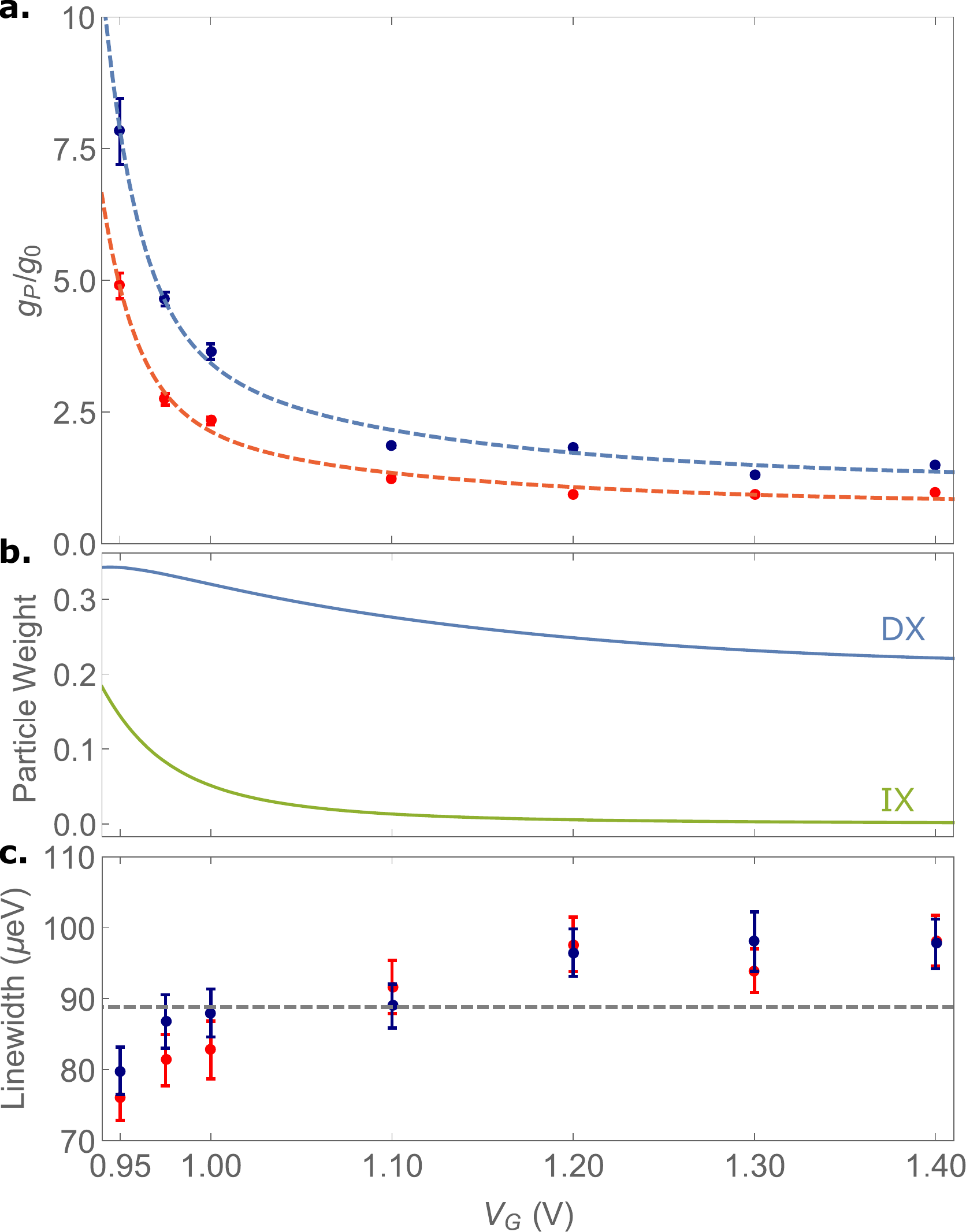}
\caption{\textbf{Increase of polariton interactions with increasing IX content.} \textbf{a.} Changes in polariton interaction strength $g_{\mathrm{P}}$  extracted from differential reflection measurements similar to Figure 2\textbf{b} as a function of $V_{\mathrm{G}}$ at a fixed position. The data match a simple model $g_{\rm P} = a \left( |x|^4 + b |y|^4 \right) g_0$ with $a = 17 \pm 1$, $b = 8 \pm 1 $ for linearly polarized measurements (red dashed line) and $a = 28 \pm 1$, $b = 8 \pm 1$ for circularly polarized measurements (blue dashed line). For each $V_{\mathrm G}$ value we repeat the experiment more than 10 times and fit a value for $g_{\mathrm P}$ to each of these measurements. The errorbars are the standard deviation of the extracted $g_{\mathrm P}$ for these measurements. \textbf{b.} Change in the particle weight as a function of $V_{\rm G}$. The dipole size of the polariton changes from 0.09 nm to 3.4 nm as $V_{\mathrm G}$ changes from 1.4 V to 0.95 V (see Supplementary Figure 9). \textbf{c.} Changes in the linewidth extracted from a reflection measurement  similar to Figure~\ref{FigSaturation}\textbf{a} for each of the points. Errorbars are the 1 $\sigma$ confidence interval of the linewidth extracted from a fit of the reflection spectrum to a Lorentzian.
}
\label{FigIXContent}
\end{figure}

We measure changes in $g_{\rm P}$ at a fixed position as we vary IX content by changing the $V_{\mathrm G}$ when the incident laser is either linearly or circularly polarized.  As Figure~\ref{FigIXContent} illustrates $g_{\mathrm P}$ increases by a factor of 5 as IX content increases from $|y|^2 = 0.002$ at $V_\mathrm{G} = 1.4 $ V to $|y|^2 = 0.14$ at $V_{\mathrm{G}} = 0.95 $ V. Remarkably, this substantial increase is independent of the polarization of the excitation laser. These results demonstrate that it is possible to enhance the nonlinearity by a factor of 5. This enhancement is accompanied with a reduction in polariton linewidth, allowing for a factor of 6.5 improvement in the ratio of nonlinearity to linewidth.

Changes in the polariton interactions are captured by a simple model that assumes $g_{\mathrm P} = a (|x|^4 + b |y|^4) g_0$. As shown in Figure~\ref{FigIXContent}\textbf{a}, measurements using linear ($a = 17$ and $b = 8$) and circularly polarized light ($a = 28$ and $b = 8$) agree with this simple model. The coefficient $b$ characterizes the ratio of IX-IX interactions to DX-DX interactions. Our measured value of $b$ is comparable to the ratio of the (direct) dipole-dipole interaction for IXs~\cite{nalitov_voltage_2014, nalitov_nonlinear_2014, kyriienko_spin_2012, byrnes_effective_2014} to the exchange based interaction for DXs~\cite{ciuti_role_1998,tassone_exciton-exciton_1999}: $\frac{e^2 d}{\epsilon \epsilon_0}\times \frac{1}{ 6 E_{\rm B} a_{\rm B}^2} \sim 30\times\frac{1}{6} = 5$, where $ \epsilon$ and $\epsilon_0$ are electric permittivities in GaAs and vacuum, $E_{\rm B}$ and $a_{\rm B}$ are the binding energy and the Bohr radius of the DX. However such an estimate does not account for the spin characteristics of excitons, induced dipole moment of the DX exciton, and the contribution of exchange to the IX interactions~\cite{nalitov_voltage_2014, nalitov_nonlinear_2014, kyriienko_spin_2012, byrnes_effective_2014}. We emphasize that our experiments, that are carried out using resonant excitation in a microcavity with coupled quantum well structures yields a much smaller increase compared to recent measurements under non-resonant excitation obtained with induced dipoles in wide quantum wells embedded in waveguide structures~\cite{rosenberg_strongly_2018}.

Our experiments unequivocally demonstrate that the interactions between IXs with a permanent dipole moment of $\sim 1000$~Debye are larger than interactions between DXs. These strong interactions when combined with the decrease in the polariton linewidth with IX content measured in Figure~\ref{FigIXContent}\textbf{b} allowed us to increase the ratio of interaction strength between dipolar polaritons to their linewidth by a factor of 6.5. A detailed analysis of our methods and results indicates that the factor of 5 increase of $g_{\mathrm P}$ is a lower bound on the enhancement of the nonlinearity (see Supplementary Notes 4 and 5). An estimate based on the blue shift of the polariton resonance yields an enhancement of $g_{\mathrm{P}}$ by a factor of 10  (see Supplementary Note 5). This significant enhancement could lead to observation of strong quantum correlations of polaritons~\cite{kyriienko_tunable_2014} when combined with state-of-the-art zero-dimensional (0D) cavities~\cite{munoz-matutano_quantum-correlated_2017,fink_signatures_2017,rodriguez_probing_2017}.  Moreover, their permanent dipole moment provides new perspectives for electrical confinement of dipolar polaritons~\cite{schinner_electrostatically_2011, winbow_electrostatic_2011, chen_artificial_2006, gartner_micropatterned_2007}: the resulting 0D polaritons would combine narrow linewidths of dipolar polaritons with confinement dimensions smaller than $\sim$ 600 nm~\cite{schinner_confinement_2013}. In addition, electrically defined lattices of 0D dipolar polaritons allow for tunability of on-site interaction strength to polariton-hopping ratio, while allowing for realization of artificial gauge fields for polaritons~\cite{lim_electrically_2017}. Even though the experiments we report are based on GaAs based heterostructures, our advances can be directly applied to van der Waals heterostructures~\cite{rivera_observation_2015,sidler_fermi_2016} to realize dipolar polaritons of transition metal dichalcogenide heterobilayers embedded in dielectric cavities.



 \paragraph*{\textbf{Acknowledgements}}
The Authors acknowledge insightful discussions with  Thomas Fink and Jacqueline Bloch. This work is supported by SIQURO, NCCR QSIT and an ERC Advanced investigator grant (POLTDES).



\end{document}